\documentclass[useAMS,usenatbib,usegraphicx]{mn2e}
\usepackage{color}

\title[Using conditional entropy to identify periodicity]{Using conditional entropy to identify periodicity}
\author[M. J. Graham et al.]{Matthew~J.~Graham,$^1$\thanks{E-mail:mjg@caltech.edu} Andrew~J.~Drake,$^1$
S.~G.~Djorgovski,$^1$   
\newauthor
Ashish~A.~Mahabal,$^1$, Ciro~Donalek$^1$\\
$^{1}$California Institute of Technology, 1200 E. California Blvd, Pasadena, CA 91125, USA}
\begin{document}

\date{Accepted . Received ; in original form}

\pagerange{\pageref{firstpage}--\pageref{lastpage}} \pubyear{2011}

\maketitle

\label{firstpage}

\begin{abstract}
This paper presents a new period finding method based on conditional entropy that is both efficient and accurate. We demonstrate its applicability on simulated and real data. We find that it has comparable performance to other information-based techniques with simulated data but is superior with real data, both for finding periods and just identifying periodic behaviour. In particular, it is robust against common aliasing issues found with other period-finding algorithms.
\end{abstract}


\section{Introduction}
The growing amount of astronomical time series data provided by the new generation of synoptic sky surveys, e.g., CRTS (\cite{crts}), PTF (\cite{ptf}), Pan-Starrs (\cite{panstarrs}), LSST (\cite{lsst}), has fostered a renewed interest in period finding algorithms, e.g., \cite{correntropy, ckp}, \cite{lasso}, \cite{fastfourier}, \cite{baluev}. There is a particular emphasis on efficiency, both in terms of speed and accuracy, to facilitate tractable analyses of tera- and petascale data sets. Period finding techniques can be divided into a number of types. The most popular seek to model a light curve via a least-squares fit to some set of (orthogonal) basis functions, most commonly trigonometric, such as Lomb-Scargle (\cite{lomb, scargle} and its derivatives/extensions (e.g., \cite{gls}), though more complicated function sets, such as wavelets (\cite{foster}), have also been tried. Another approach is to minimize some measure of the dispersion of time series data in phase space, such as binned means (\cite{stellingwerf}), variance (\cite{aov89}) or entropy (\cite{cnm95}), which can often be regarded as an expansion in terms of periodic orthogonal step functions. Bayesian methods (\cite{gregory92}, \cite{wang12}) are also becoming common and  
there have even been attempts to search for periodicity using neural networks (\cite{baluev12}).

The basis of an algorithm also often determines how well it copes with the real world aspects of time series data, such as irregular sampling, gaps, and errors, e.g., standard Fourier analysis is impossible for any data diverging from regular sampling. \cite{dejager} argue that in the case of weak signals, most period finding methods only work well with certain kinds of periodic shapes and that this causes a selection effect for the general identification of weak periodic signals. Similar shape dependencies are found in \cite{schwarzenberg99}.

Intuitively the fastest period finding algorithm will involve a single pass through a data set per trial period and integer counting operations, e.g., histogram binning. Any higher-order function calls, particularly per data point in a time series, will extend the average calculation time per trial period and, consequently, the overall time taken by the algorithm to determine a correct period. 

Among the different types of approach -- Fourier-based, Bayesian, autoregressive modelling, etc., one of the most promising is information theory as these type of techniques seem better equipped to deal with uneven sampled time series (as most modern light curves are). Information theory-based methods extract information from the probability density function and so include higher-order statistical moments present in the data whereas Fourier  or analysis of variance techniques are based only on second-order statistical analyses. This implies that information theory brings better modelling of the underlying process and robustness to noise and outliers. \cite{ckp} employ information theory-based statistical descriptors, such as Renyi quadratic entropy and correntropy, which are generalizations of second order moment statistics such as variance and correlation. 

\cite{cnm95} introduced a method to find the period of an (irregular-sampled) time series by minimizing its Shannon entropy when folded by a trial period. The idea is that a light curve folded at most trial periods will produce a random arrangement of points in a particular region, a unit square, say, whereas, when folded at the correct period, the light curve will be the most ordered arrangement of data points in the region and so contain the most information about the signal. As entropy measures the lack of information about a system, the correct period minimizes this quantity. Moreover, this can be formally proven to be mathematically correct within the framework of information theory (\cite{chmn99}) whilst other measures based on the statistical analysis of the ``shape'' of the light curve lack a formal proof. 

In this work, we introduce a new technique based on the {\em conditional} Shannon entropy of a light curve.
This has the advantage of accounting for systematic effects in the phase space coverage of time series, i.e., 
gaps, concentrations, and other artifacts that may be present in the phase distribution when the light curve is folded by a trial period as a result of sampling, etc. 

The paper is structured as follows: in section 2, we present the new algorithm and in section 3, the data sets we have applied it to. We discuss our results in section 4 and conclusions in section 5.

\section{Algorithms}

\subsection{Conditional entropy}

Formally, a time series, $m(t_i)$, is normalized to occupy a unit square in the $(\phi, m)$ plane where $\phi_i$ is the phase at $t_i$ given a trial period, $p$, such that $\phi_i = t_i / p - [t_i / p]$, where the square brackets denote the integer function. The unit square is then partitioned into $k$ partitions (bins) and the (Shannon) entropy for the distribution, $H_0$, defined by: 

\begin{equation}
H_0 = - \sum^{k}_{i = 1} \mu_i \ln(\mu_i); \forall \mu_i \ne 0
\end{equation}

\noindent
where $\mu_i$ is the occupation probability for the $i^{th}$ partition, which is just the number of data points in that partition divided by the total number of points in the data set.

\begin{figure*}
\caption{This shows the light curve of a typical type AB RR Lyrae from CRTS (Drake et al. 2013) (a) folded at the trial period which minimizes the entropy (b) and conditional entropy (c).}
\label{solar}
\begin{tabular}{lll}
(a) & (b) & (c) \\
\includegraphics[width=2.1in]{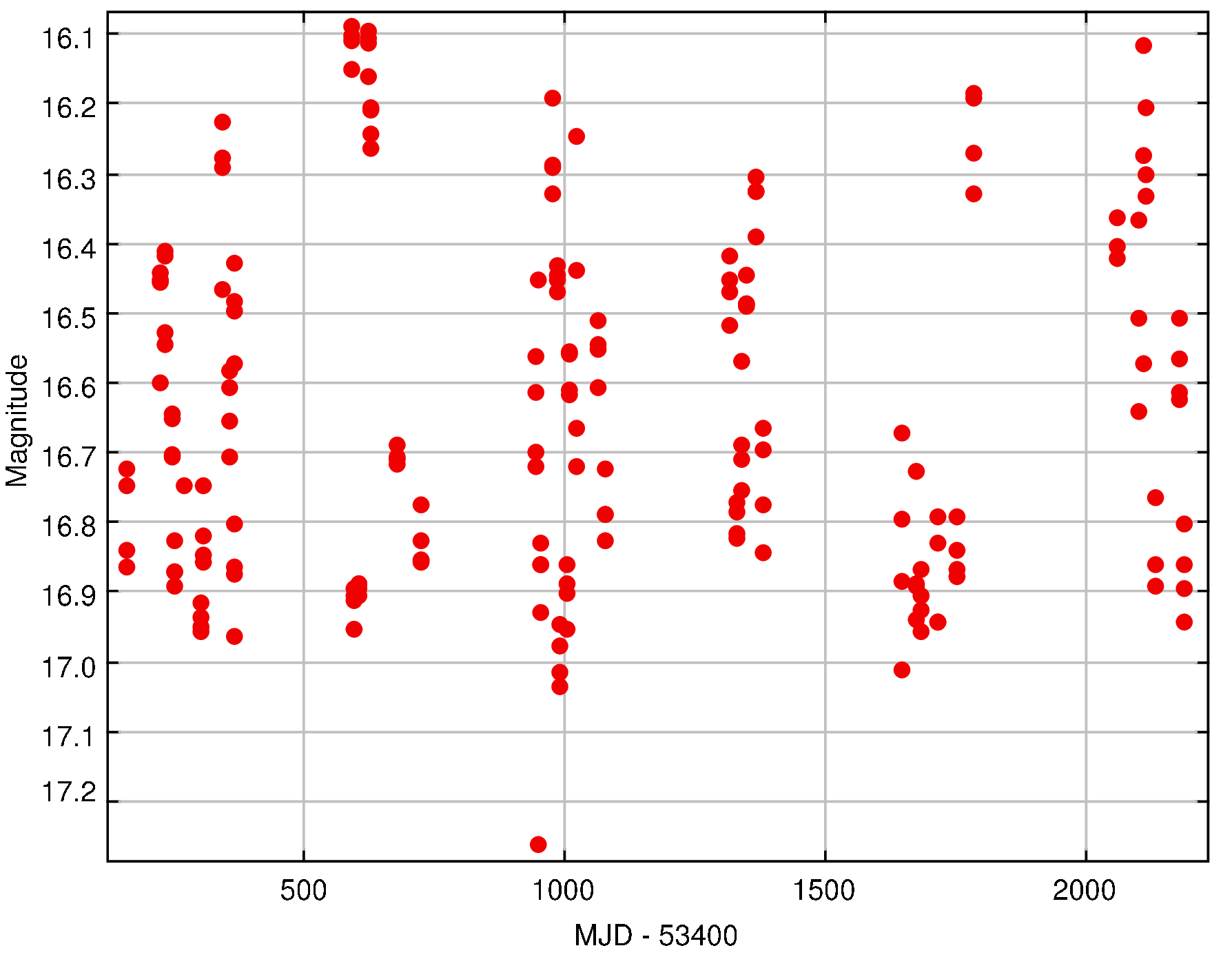} &
\includegraphics[width=2.1in]{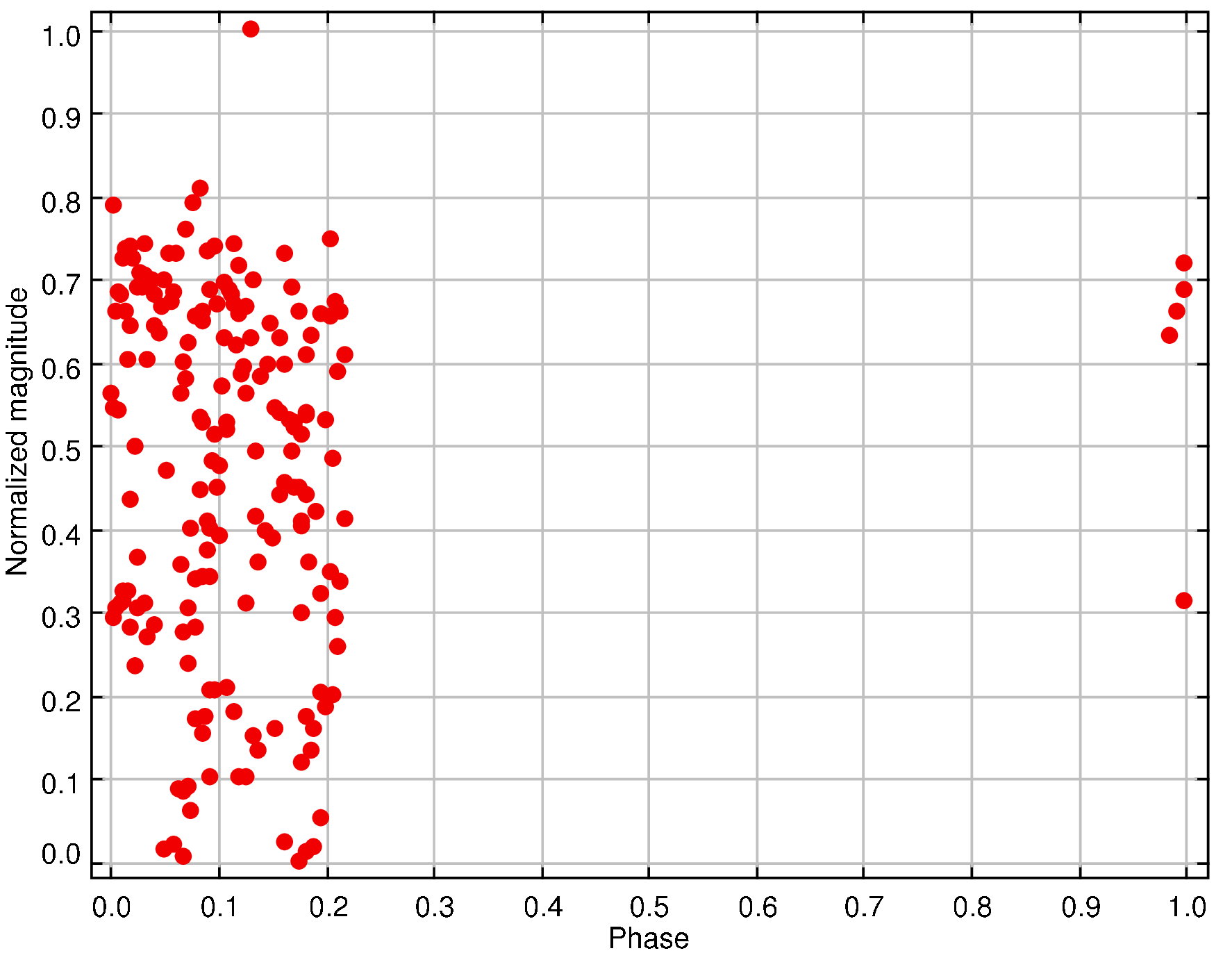} & 
\includegraphics[width=2.1in]{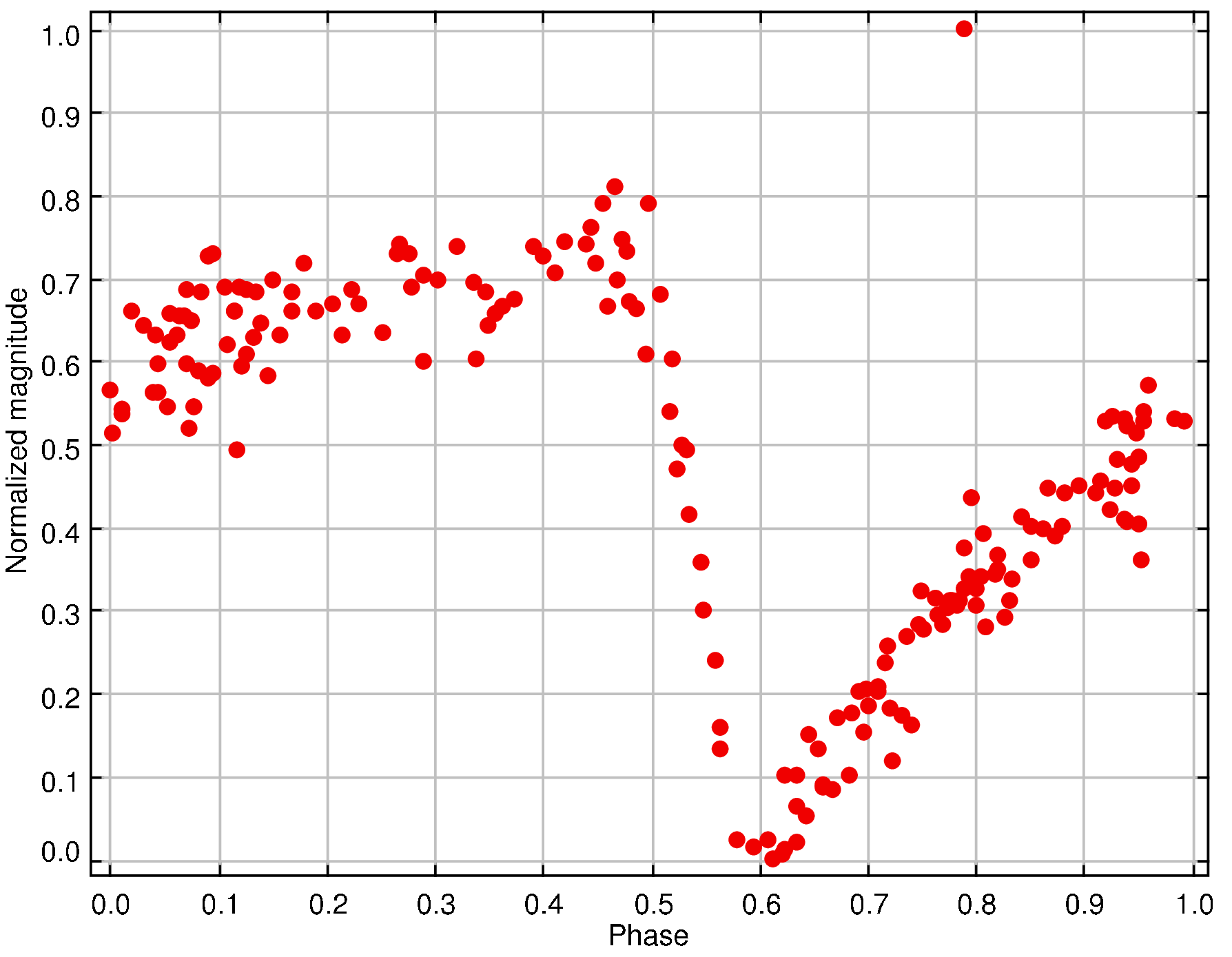} \\
\end{tabular}
\end{figure*}

However, on applying this method to real data, e.g., a typical type AB RR Lyrae from CRTS (\cite{rrlyrae}) (see Fig.~\ref{solar}), we found that the period which minimized the entropy was predominantly that associated with the mean solar day (p = 1.00274 days). Looking at a folded light curve at this period (see Fig.~\ref{solar}(b)), it is clear that this does indeed produce the most ordered arrangement of points in terms of compactness of points within the unit square; however, this is not the most ordered in terms of an underlying functional support which the correct period would produce. Another way of expressing this is that with the solar period, the order of points per phase interval is not optimized whereas it is with the true period - the amount of randomness in the normalized magnitude is minimized given the known values of the phase. We note that this effect can be mitigated to some degree through an appropriate choice of partition (\cite{cincotta99})but this then introduces an additional step into the period finding process.

A related quantity taking this into account is the {\em conditional} entropy, $H(m|\phi)$, defined by:

\begin{equation}
H_c =  \sum_{i,j} p(m_i, \phi_j) \ln \left( \frac{p(\phi_j)}{p(m_i, \phi_j)} \right)
\end{equation}

\noindent 
where $p(m_i, \phi_j)$ is the occupation probability for the $i^{th}$ partition in normalized magnitude and the $j^{th}$
partition in phase and $p(\phi_j)$ is the occupation probability of the $j^{th}$ phase partition, which for rectangular partitions is just:

\[
p(\phi_j) = \sum_i p(m_i, \phi_j)
\]

\begin{figure}
\caption{This shows the conditional entropy periodogram (frequency in days$^{-1}$) for the light curve of a CRTS RRAB in Fig.~\ref{solar}. Note that there is no discernible minimum at the mean solar day period (1.00274 d).}
\label{periodogram}
\includegraphics[width=3.2in]{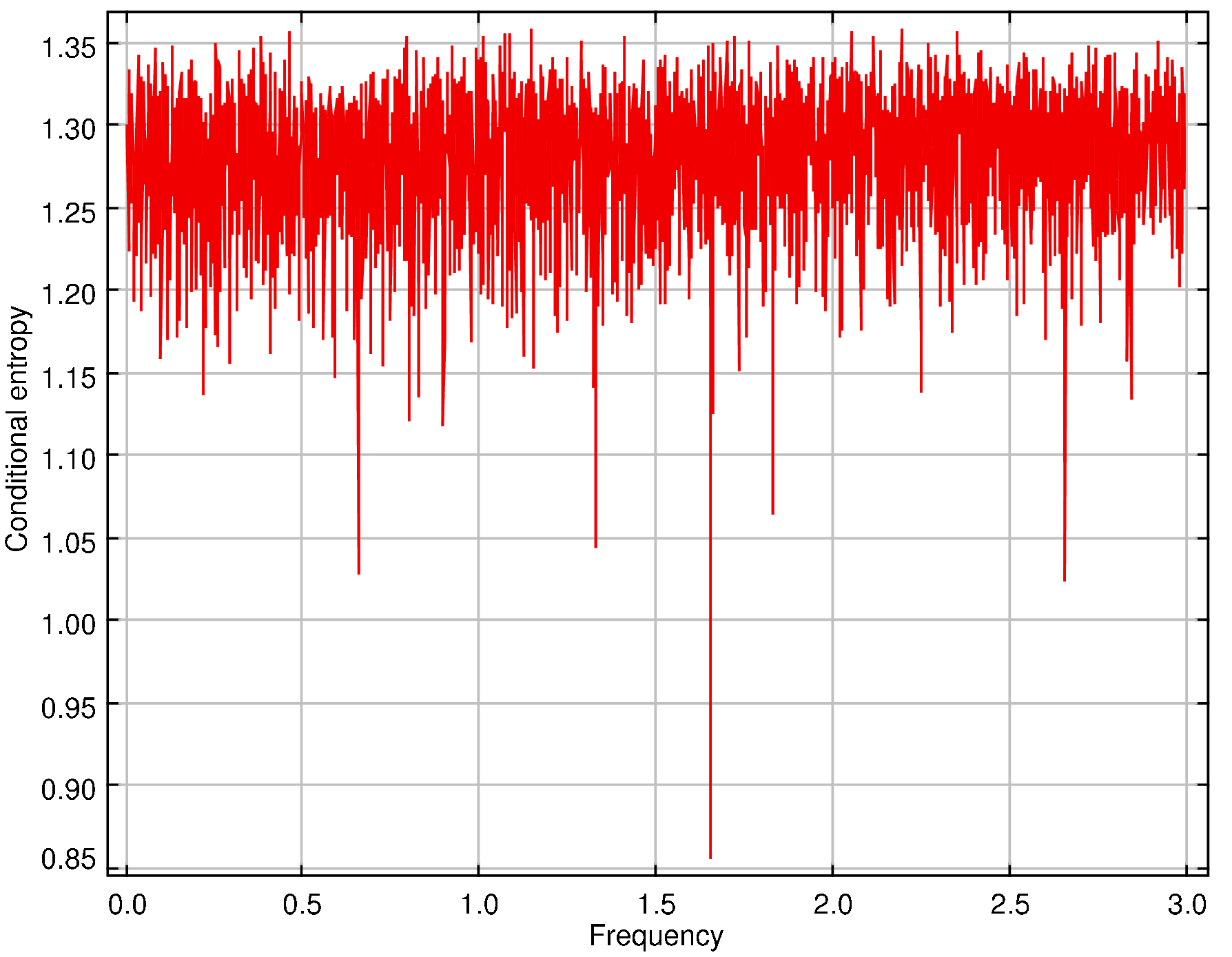}
\end{figure}

Since the definition of $H_c$ is not dependent on the partition shape, we also consider an optimal estimator for $H_c$ based on an optimal partitioning of the data using Bayesian blocks (\cite{bb}) (see Appendix~\ref{appa} for details). We found good agreement between the values of the conditional entropy for the two partition schemes: applying the two estimators to the same data set produces conditional entropies that are strongly correlated. Differences in numerical value are attributable to the lack of normalization in either estimator.
However, the optimal partitioning scheme is computationally expensive since it involves determining the Voronoi tessellation for each trial period and this precludes it from being an efficient period finding algorithm. 
We therefore adopt a simple rectangular partitioning scheme in this analysis.

Fig.~\ref{solar}(c) shows the light curve folded at the trial period which minimizes the conditional entropy (see Fig.~\ref{periodogram} for the associated periodogram - the plot of test statistic vs. frequency. The search for the correct period is most commonly done using a scan through frequency space).

The most likely periods are those associated with the strongest minima in the periodogram. Although it is possible to associate a probabilistic significance with a particular value (\cite{chmn99}) of this statistic, this is not a powerful enough discriminator between rare (likely) events. Rather we decided to see whether using an additional statistic to identify the correct period from the subset identified by the CE algorithm could boost the overall performance.

Thus to assess which is the most significant of the CE minima, we calculate an analysis-of-variance (AOV; \cite{aov89}) statistic for the $k$ most likely periods and select the period which maximizes this statistic as the measured period for this time series. 

We observe as well that, in some cases, the CE periodogram is not flat but exhibits a weak overall dependency on frequency, i.e., there is a trend for generally lower CE values at lower frequencies. This could lead to misidentification of the strongest minima in the periodogram and so we add a normalization step of dividing the periodogram by a smoothed version using a wide rolling median filter before identifying the strongest minima.

Finally, we note that, as with the original Shannon entropy-based method of \cite{cnm95}, the algorithm does not yet explicitly take into account errors on the data. \cite{cincotta99} addressed this with essentially a kernel-based estimator for the Shannon entropy and an equivalent expression is easily derivable for the conditional entropy. It is less efficient, though, as simple integer counting operations have been replaced with more complicated function calls. The effect of errors in the data are also somewhat mitigated by our use of overlapping partitions (see below) with individual data points contributing to the occupation probabilities of more than one bin as they would with a kernel.

\subsection{Period harmonics}

One particular issue for automated period finders (particularly Lomb-Scargle) is that they misidentify a multiple of the period as the ``true'' period - this is a common problem for binary systems where the half period is frequently the most significant peak in a periodogram. For example, \cite{macc12} initially find 70\% of their periods for eclipsing binaries (EBs; $\sim$49\% of all objects) in the ASAS Catalog of Variable Stars (ACVS; \cite{acvs})  to be half periods. As discussed in \cite{wang12}, this is attributable to two aspects: for symmetric EBs, the true period and half its value are not clearly distinguishable quantitatively. Meanwhile, methods that are successful for EBs tend to find integer multiple periods of ``single bump" stellar types, such as RR Lyrae and Cepheids, and vice versa.

Several techniques have been proposed to deal with this. \cite{pdm2} suggests ``subharmonic averaging'' where a significant signal in the periodogram (test statistic vs. frequency) is replaced by the average of the statistic value at the peak frequency (that associated with the significant statistic value) and its value at half the peak frequency. For real signals, the statistic value will be boosted whilst for false signals, the statistic value will decrease significantly. This can be computationally expensive, however, as it involves scanning through all the trial frequencies (periods) used. \cite{wang12} propose including domain knowledge via a probabilistic generative filter that attempts to match light curves, folded at both the best identified periods and their doubled values, to the learned shapes of common object types with the most likely giving the assumed value. Use of the filter gives an 18\% improvement in the accuracy of calculated periods against their quoted value. \cite{macc12} train a random forest-based supervised classifier to detect and correct for this artifact giving a 24\% boost to their accuracy, although they still find that 15.6\% of their calculated periods for all variable stars in the ACVS are actually half (14.1\%) or double (1.6\%) the true (quoted) value.

We propose a simpler approach based on fitting the light curve, $y(\phi)$, phased at a period, $p$, with a smoothing spline, $f$, which minimizes:

\[
\sum_{i = 1}^{n} \left[ \frac{y_i - f(\phi_i)}{w_i} \right]^{2} + \rho \int_{-\infty}^{\infty} (f^{\prime\prime})^{2} d\phi
\]

\noindent
where $w_i$ are the relative weights for each point and $\rho$ is a smoothing parameter determined by a generalized cross-validation technique (\cite{gcvs85}). Note that $f$ is necessarily a natural cubic spline with knots at $\phi_i$ for $i = 1, \ldots, n$. We identify the strongest dip (minimum) in the spline and then repeat the procedure for the light curve phased at double the period, i.e. $2p$, and find the two strongest dips there. For an object where the measured period is the true period, $p = p_0$, the two dips in the $2p$-spline should be of the same amplitude within some measurement tolerance and also the same as the dip in the $p$-spline; however, for an object with $p = p_0 / 2$, i.e., most likely an eclipsing source, there should be a discernible difference between the two dips in the $2p$-spline, although this will not be generally true for the subclass of binaries which have equivalent minima, i.e., W UMa-type variables. We therefore consider the doubled period as the true value for objects where the difference between the two dips is greater than some threshold, the photometric error for the light curve, say, and the difference between the smaller of the two dips in $2p$-spline and the dip in the $p$-spline is also greater than a similar threshold. We note, though, that this threshold value may also be dependent on the signal-to-noise ratios of light curves within a particular survey.

\subsection{Data binning}

Many period finding algorithms use data binning (normally just of the phase (folded period) variable) in calculating their test statistic. The choice of binning parameters - width, number and location - can therefore have a significant effect on the resolving power of a particular method: too wide a bin leads to folded curves with similar phase distributions having the same test statistic, whilst too narrow a bin means that the test statistic is dominated by small number contributions giving a noisy representation of the phase distribution. \cite{kovacs} describes a process for the optimal phase cell number of a phase dispersion measure statistic that depends on the data length, signal form and noise level. There are also a number of more general prescriptions for selecting the optimal binning parameters when binning data - Bayesian blocks mentioned previously and jackknife likelihood (\cite{hogg08}) - or replacing the binning entirely with a suitable Bayesian prior (\cite{loredo11}). There is, however, no overall optimal approach amongst these.

In a sweep through a frequency (period) range, the phase distribution will vary as the trial frequency (period) varies and thus the optimal bin widths and number of bins required to cover it. However, it is computationally expensive to calculate these optimal values for each specific trial frequency and so fixed ``mean'' optimal values are used in the relevant algorithms. We have determined the range of the optimal number of bins and bin widths for a set of sample light curves with numbers of observations spanning the range $\sim 10 - 2000$ using both the jackknife and Bayesian block approaches. We find that a phase bin width of $\Delta \phi = 0.1$ (giving 10 bins) is close to optimal and use this for the algorithm; we also use a magnitude bin width of $\Delta m = 0.2$, determined in a similar fashion.

AOV makes use of flexible bin sizes when there is poor phase coverage and less than 5 points in some bins. We have adopted a similar approach for our algorithm, using an overlapping bin of width $\Delta \phi = 0.2$ to calculate $H_c$ and accounting for data points being included twice, e.g., a point at $\phi = 0.25$ will be included in both the bins covering $\phi = 0.1 - 0.3$ and $\phi = 0.2 - 0.4$ respectively, when there is poor phase coverage. The PDM2 algorithm (\cite{pdm2}) also follows a similar strategy.

We also omit all points in a light curve which are defined as outliers according to:

\[
\frac{|x_i - \mathrm{med}_j x_j|}{\mathrm{MAD}_n} > 3.0
\]

\noindent 
where med$_i x_i$ is the sample median and MAD$_n$ is the median absolute deviation from the median.

\section{Data sets}
In this analysis, we consider synthetic data and real data from the MACHO surveys.

\subsection{Synthetic data}
We generate synthetic time series with the form:

\[
m(t) = A_0 + \sum_{n=1}^{3}A_n \sin \left( \frac{2n\pi t}{P} \right) + B \sigma
\]

\noindent 
where $A_0 = 15$, $A_1 = -0.5$, $A_2 = 0.15$, $A_3 = -0.05$, $P$ is the period, $B$ is a scaling factor ranging from 0.1 to 1.0 and $\sigma$ is a Gaussian distributed random variable with zero mean and unit standard deviation (${\cal N}(0,1)$). Periods are generated according to $P=10^{(p - 1)}$, where $p$ is a random variable drawn from a lognormal distribution with zero mean and a standard deviation of 0.75 -- this broadly mimics the stellar period distribution from variable surveys. We note that this form of synthetic data is fairly standard (e.g., \cite{cnm95}, \cite{correntropy}), apart from the scalable noise term we are employing.

\begin{figure*}
\caption{This shows sample synthetically generated time series for: (a) 100 points over 250 days with $B=0.2$ and a period of 0.752d; (b) 250 points over 1500 days with $B=0.6$ and a period of 7.52d; and (c) 500 points over 3000 days with $B=1.0$ and a period of 17.52d. }
\label{samplelc}
\begin{tabular}{lll}
(a) & (b) & (c) \\
\includegraphics[width=2.2in]{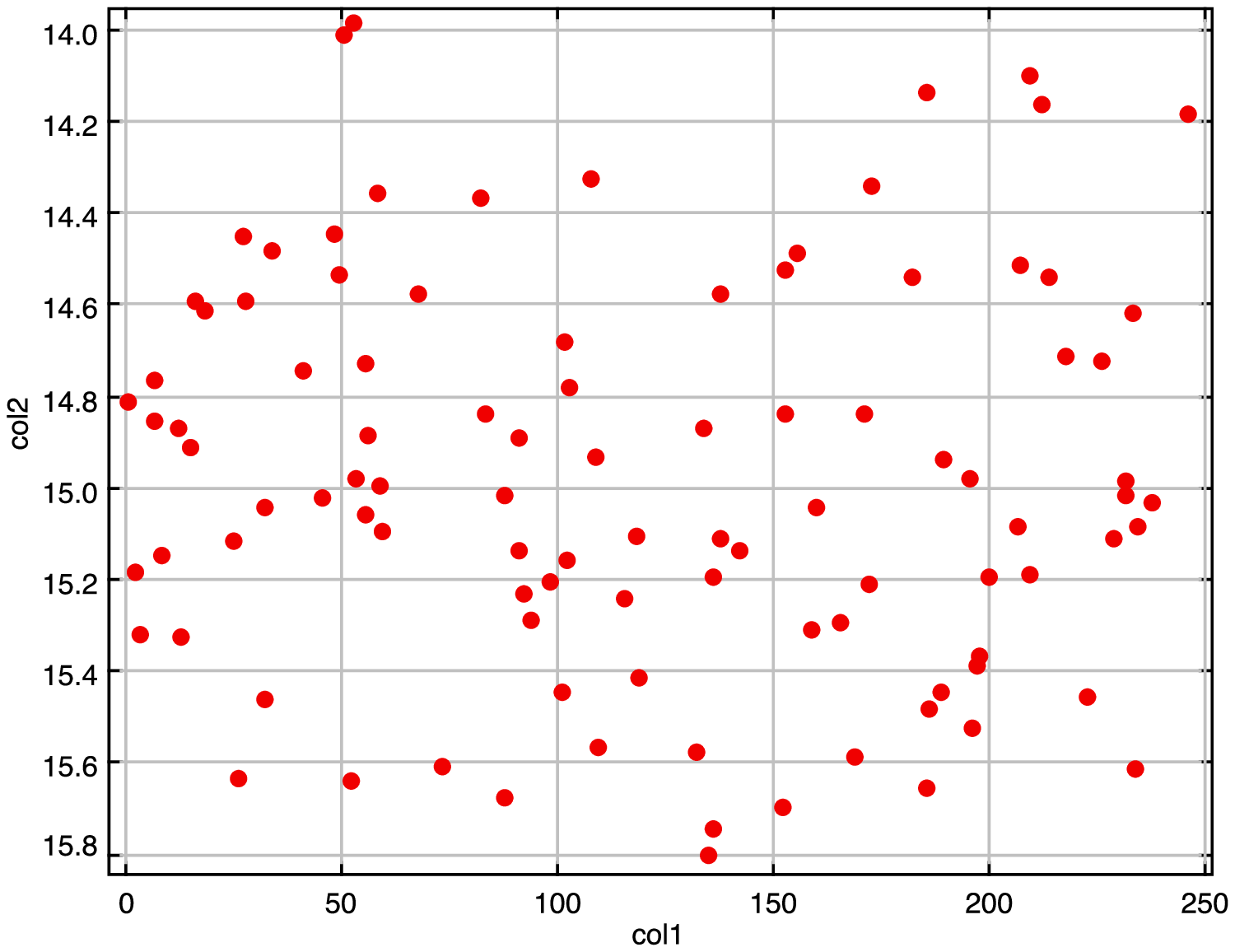} &
\includegraphics[width=2.2in]{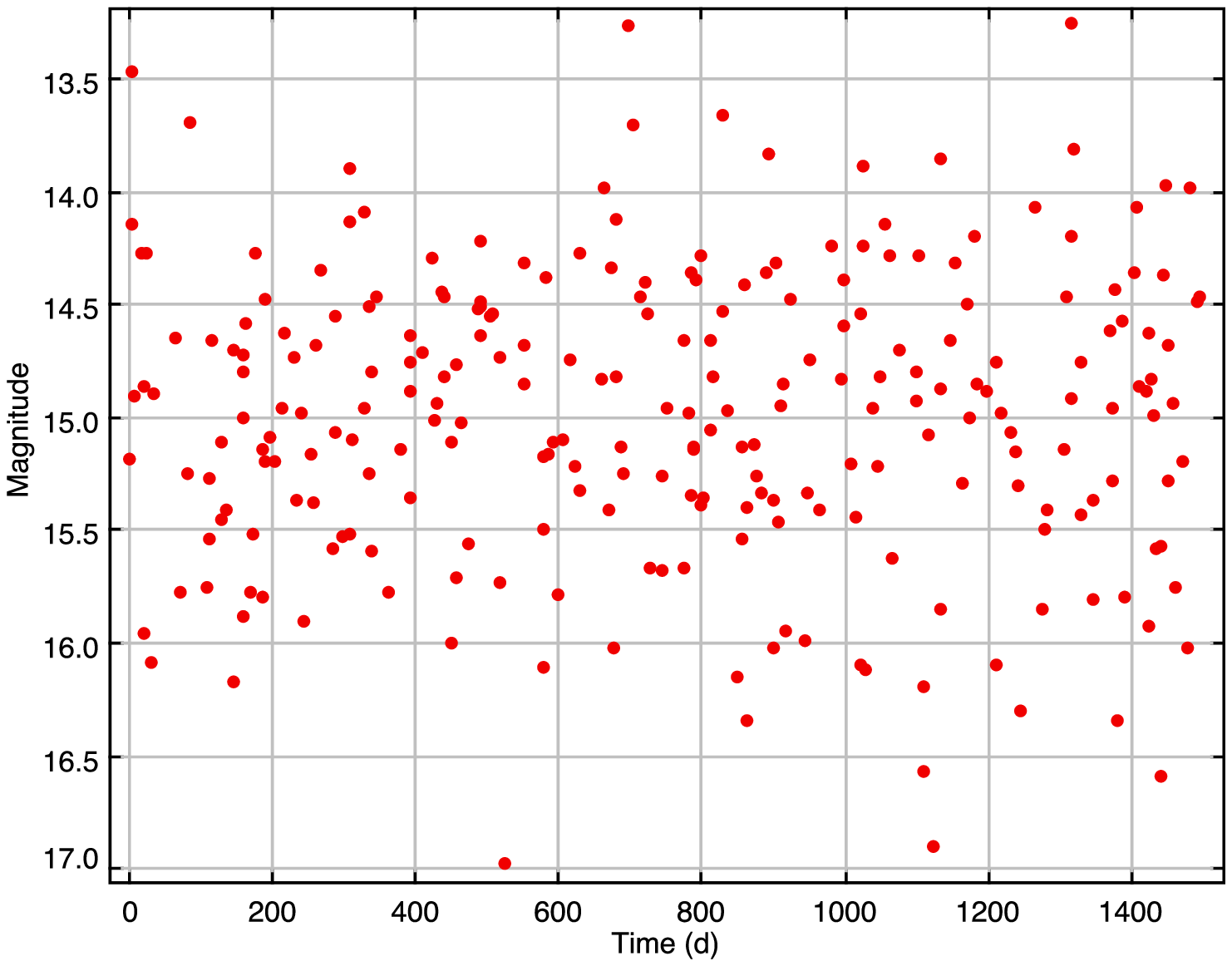} &
\includegraphics[width=2.2in]{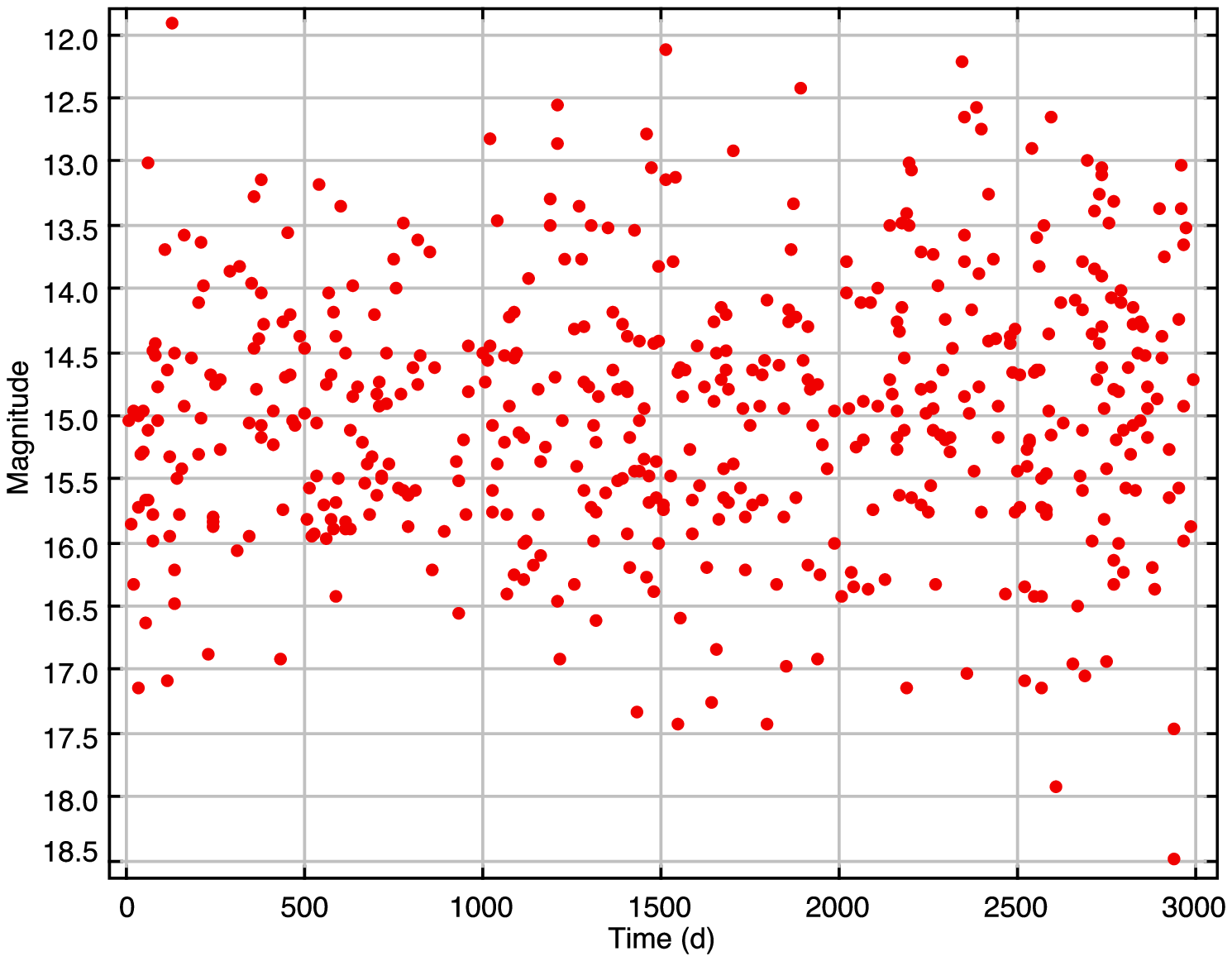} 
\end{tabular}
\end{figure*}

We have produced sets of 1000 light curves consisting of $n$ points randomly spanning a temporal baseline of $\tau$ days with noise scale $B$ for a grid of $(n, \tau, B)$, such that $n = 50$ -- $500$ with $\Delta n = 50$, $\tau = 250$ -- $3000$ with $\Delta \tau = 250$, and $B = 0.1$ -- $1.0$ with $\Delta B = 0.1$. Sample light curves are shown in Fig.~\ref{samplelc}. 

\subsection{MACHO}

The MACHO survey (\cite{macho}) was designed to search for gravitational microlensing events in the Magellanic Clouds and the Galactic Bulge and more than 20 million stars were observed, making it an important resource for variable star studies. A ``gold standard'' data set of light curves has been produced
 from the MACHO survey by the Harvard Time Series Center, consisting of approximately 500 each of RR Lyrae, eclipsing binaries and Cepheids respectively covering the LMC ($75^{\circ} < \mathrm{RA
} < 85^{\circ}$, $-71^{\circ} < \mathrm{Dec} < -67^{\circ}$). Although MACHO data normally consists of blue and red channel data for each stellar object, only the blue channel (V-band equivalent) have been used here. This data set has also been used in two correntropy-based (generalized correlation) approaches for estimating periods in non-uniformly sampled time series (\cite{misha11}, \cite{ckp}).

\section{Results}

For each of the synthetic data sets, we have estimated the efficiency of the algorithm as a function of accuracy, i.e., what fraction of 1000 light curves with different numbers of data points, temporal coverage, and noise levels does the method recover the true period to a prescribed level of accuracy. We define our accuracy in terms of the absolute difference between the recovered period and the true period relative to the true period:

 \[ 
\mathrm{accuracy} =  \frac{|P_{rec} - P_{true}|}{P_{true}}
 \]

\noindent
As we noted in section 2, a period-finding algorithm may also frequently find a period (sub)harmonic instead of the true period. To determine how close the found period is to an integer (sub)multiple of the true period, we use:

\[ 
\mathrm{accuracy} = \left| \frac{P_{rec}}{P_{true}} - \left\|\frac{P_{rec}}{P_{true}}\right\| \, \right|  \; \mathrm{for} \; P_{rec} > P_{true} 
\]

\noindent
and

\[ 
\mathrm{accuracy} = \left| \frac{P_{true}}{P_{rec}} - \left\|\frac{P_{true}}{P_{rec}}\right\| \, \right|   \; \mathrm{for} \; P_{rec} < P_{true} 
\]

\noindent
where $\|x\|$ is the nearest integer to $x$. As a comparison for the performance of the conditional entropy method, we have also tested the straightforward (Shannon) entropy algorithm of \cite{cnm95}.

For each simulated light curve with a period $P$ and $n$ observations spanning a baseline of $\tau$ days, we can determine the number of observations per cycle, i.e., the density of points in the folded light curve, and this allows us to easily compare the accuracies across our simulation grid, for example, that of objects with a period of 0.5 day and 50 observations over a 1 year baseline with those with a period of 500 days and 500 observations over a 10 year baseline.

\begin{figure*}
\caption{This shows the distribution of accuracies from the synthetic data in terms of the number of observations per cycle for the two entropy-based methods: (a) conditional entropy and (b) Shannon entropy.  The concentrations at poor accuracy and high observations per cycle originate with the noisiest simulated data (B $> 0.8$). Both methods are successful, although the conditional entropy is marginally better - it returns slightly more objects at higher accuracies.}
\label{npt}
\begin{tabular}{ll}
(a) & (b) \\
\includegraphics[width=3.45in]{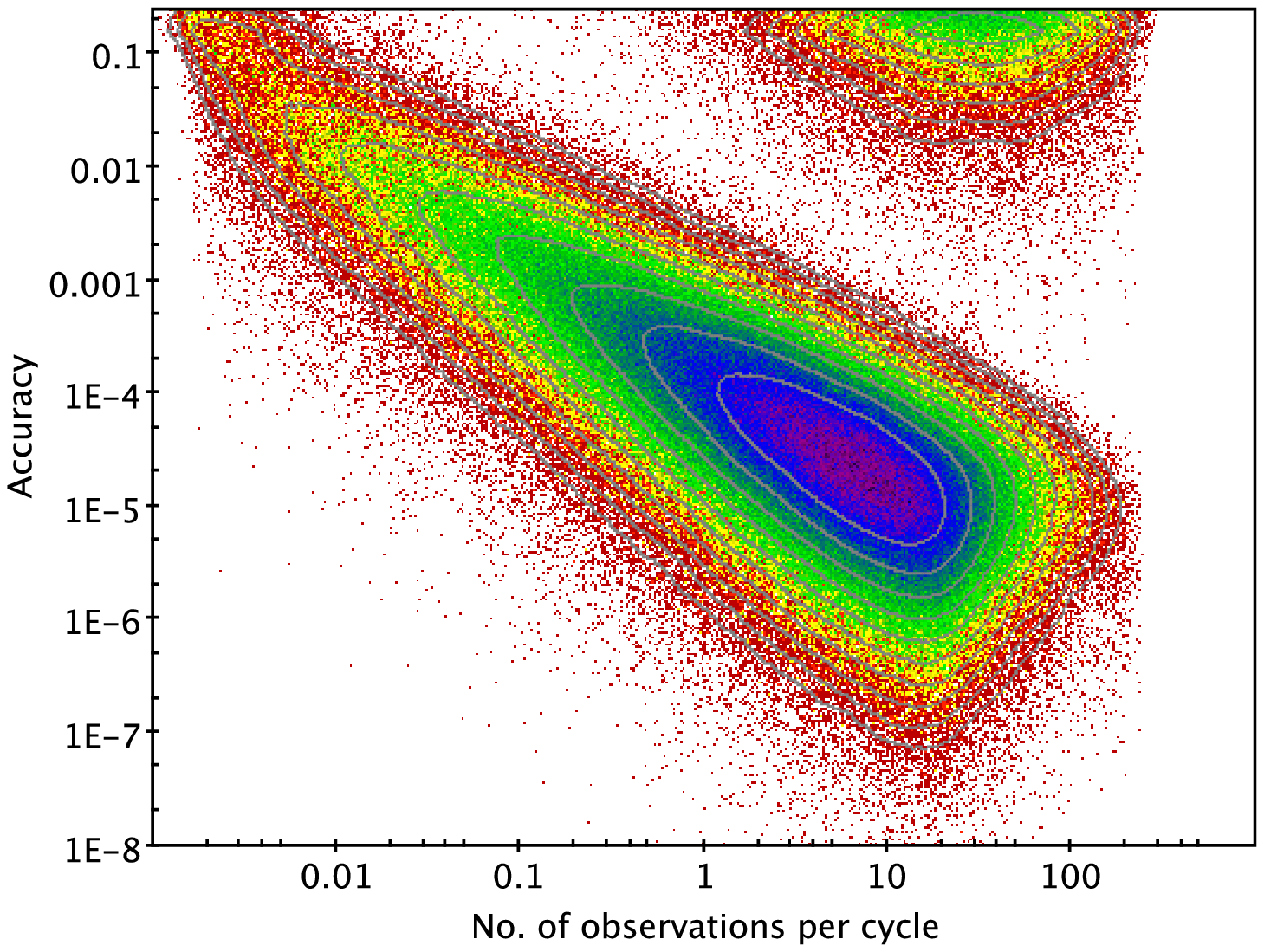} & 
\includegraphics[width=3.45in]{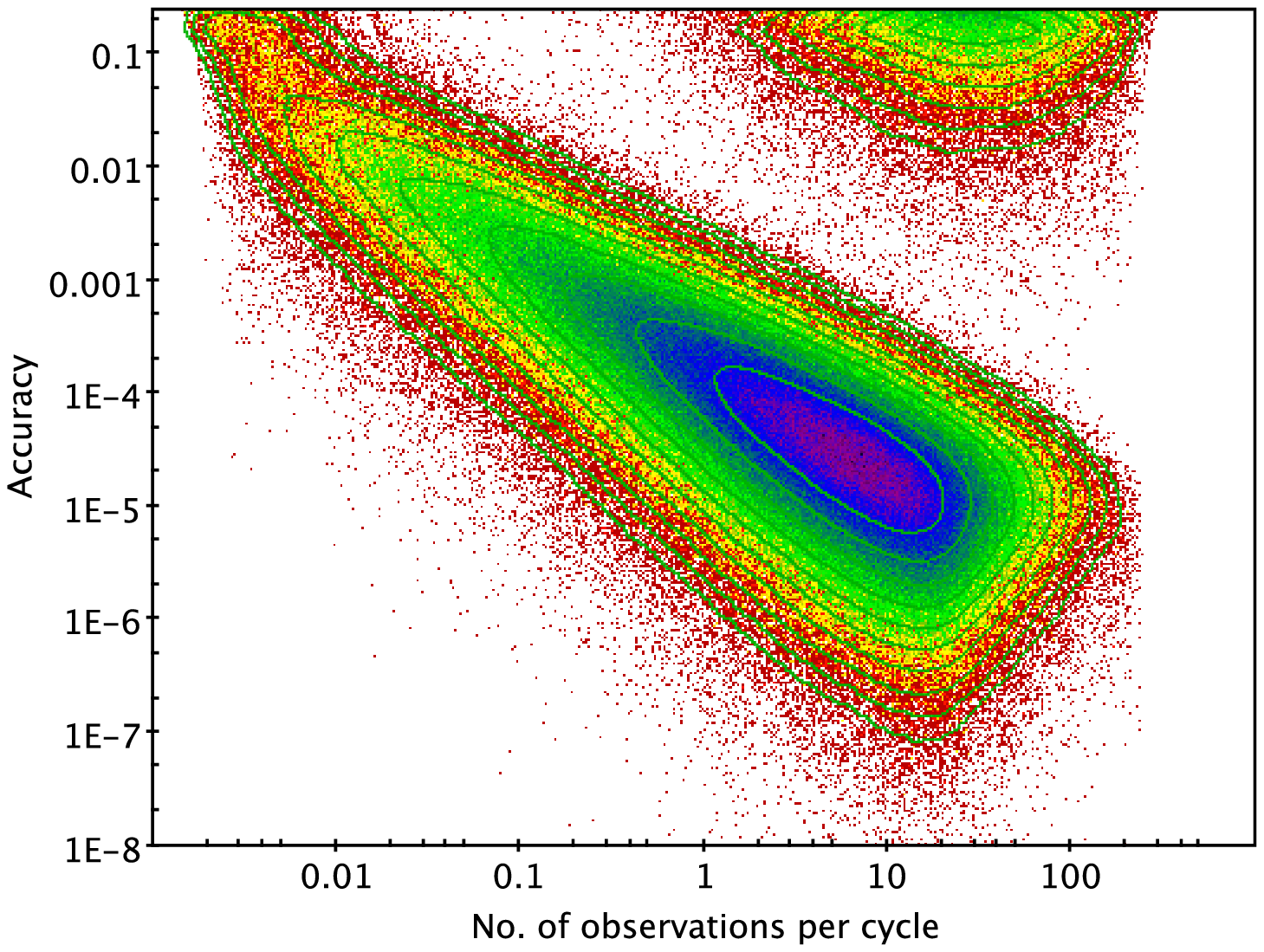} \\
\end{tabular}
\end{figure*}

\begin{figure}
\caption{This shows the tracks of the median centroids of the accuracy distributions from the synthetic data for the two entropy-based methods - red (conditional entropy) and blue (Shannon entropy) -- with the different values of the error scaling factor, $B$. }
\label{centroid}
\includegraphics[width=3.45in]{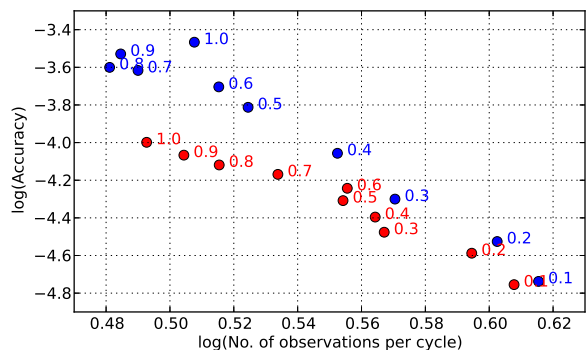}
\end{figure}

Fig.~\ref{npt} shows the distribution of accuracies against the number of observations per cycle for the two entropy-based methods with the synthetic data. Clearly the better sampled the folded light curve is, the better the accuracy of both methods, although the conditional entropy method returns a slightly higher proportion of accurate results than the regular entropy -- 5\% more of objects have an accuracy less than a $10^{-5}$ cutoff with conditional entropy than with Shannon. The tracks of the median centroid of the distributions with varying $B$ are shown in Fig.~\ref{centroid} indicating that as the light curves get noisier, both methods also get less accurate but that the Shannon method does so at a quicker rate - past $B = 0.5$ there is 0.5 dex difference in the median accuracy for the two. 

\begin{figure*}
\caption{The upper plots show the normalized distribution of accuracies of the recovered period relative to the true period for the two entropy-based algorithms for the difference values of the error scaling factor, $B$, in the synthetic data: (a) is the conditional entropy, (b) is the Shannon entropy. The lower plots show the normalized distribution of accuracies of the recovered period relative to an integer (sub)multiple of the true period: (c) is the conditional entropy, (d) is the Shannon entropy. The conditional entropy performs moderately better at higher noise levels, particularly in detecting period harmonics.}
\label{err}
\begin{tabular}{ll}
(a) & (b) \\
\includegraphics[width=3.45in]{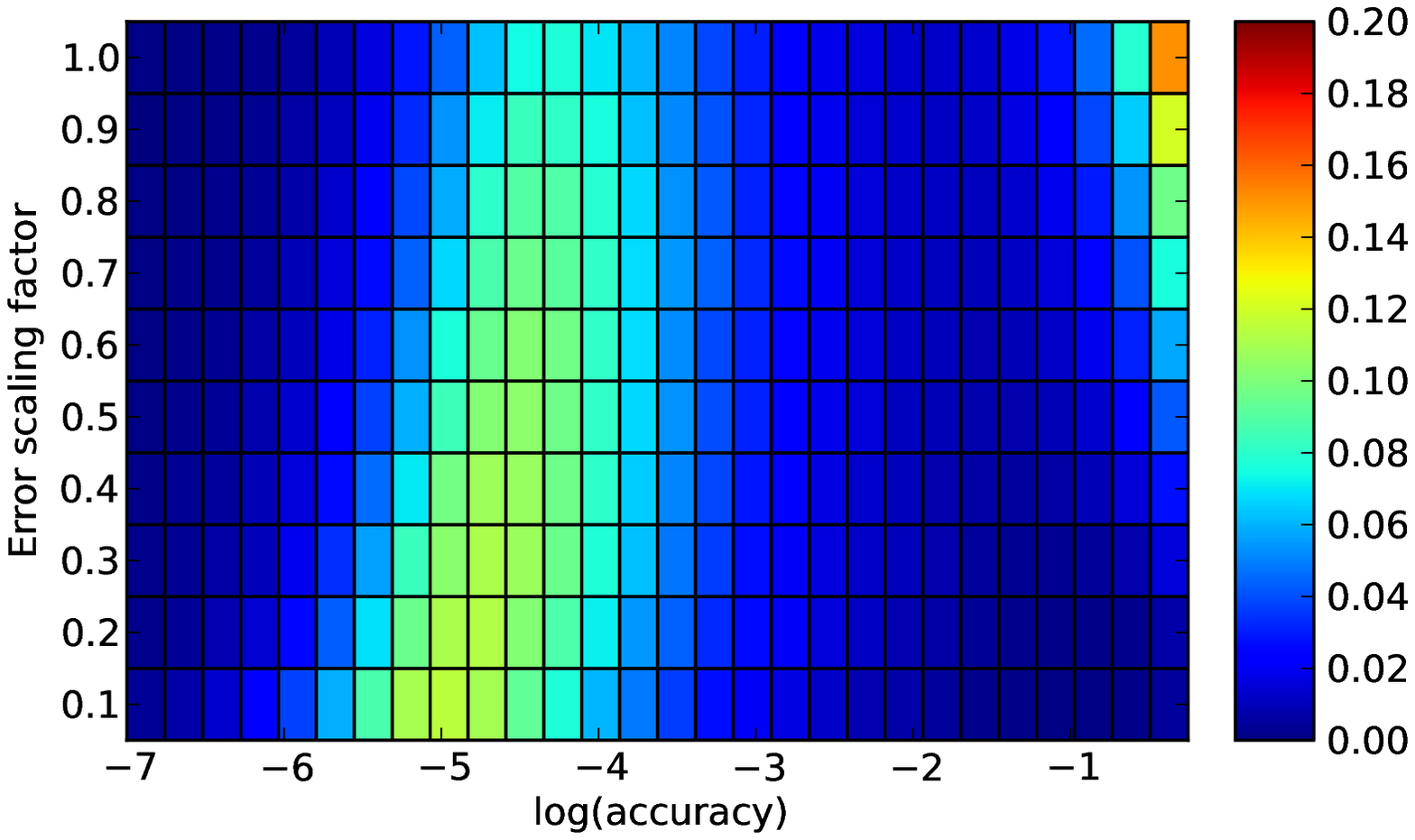} & 
\includegraphics[width=3.45in]{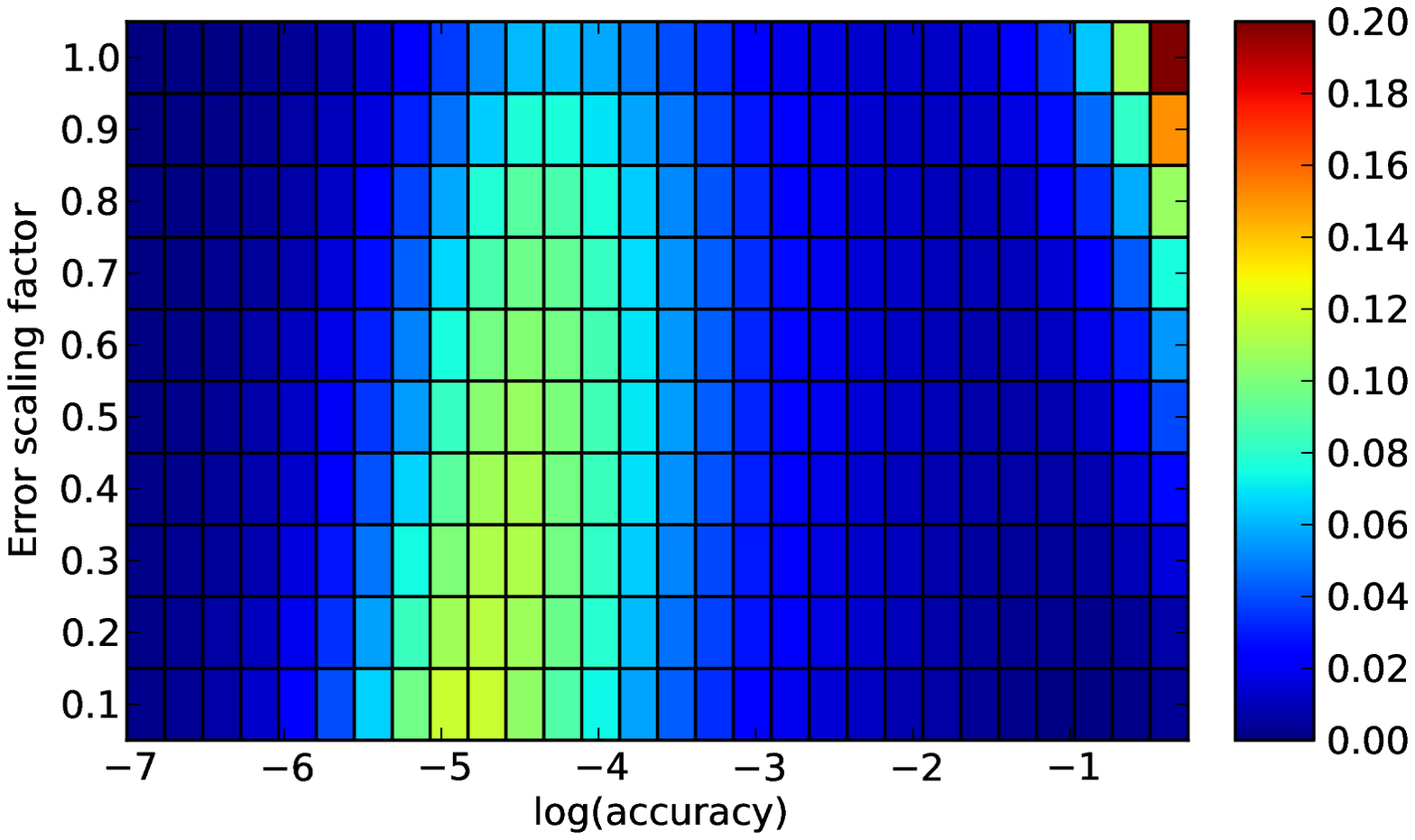} \\
(c) &  (d) \\
\includegraphics[width=3.45in]{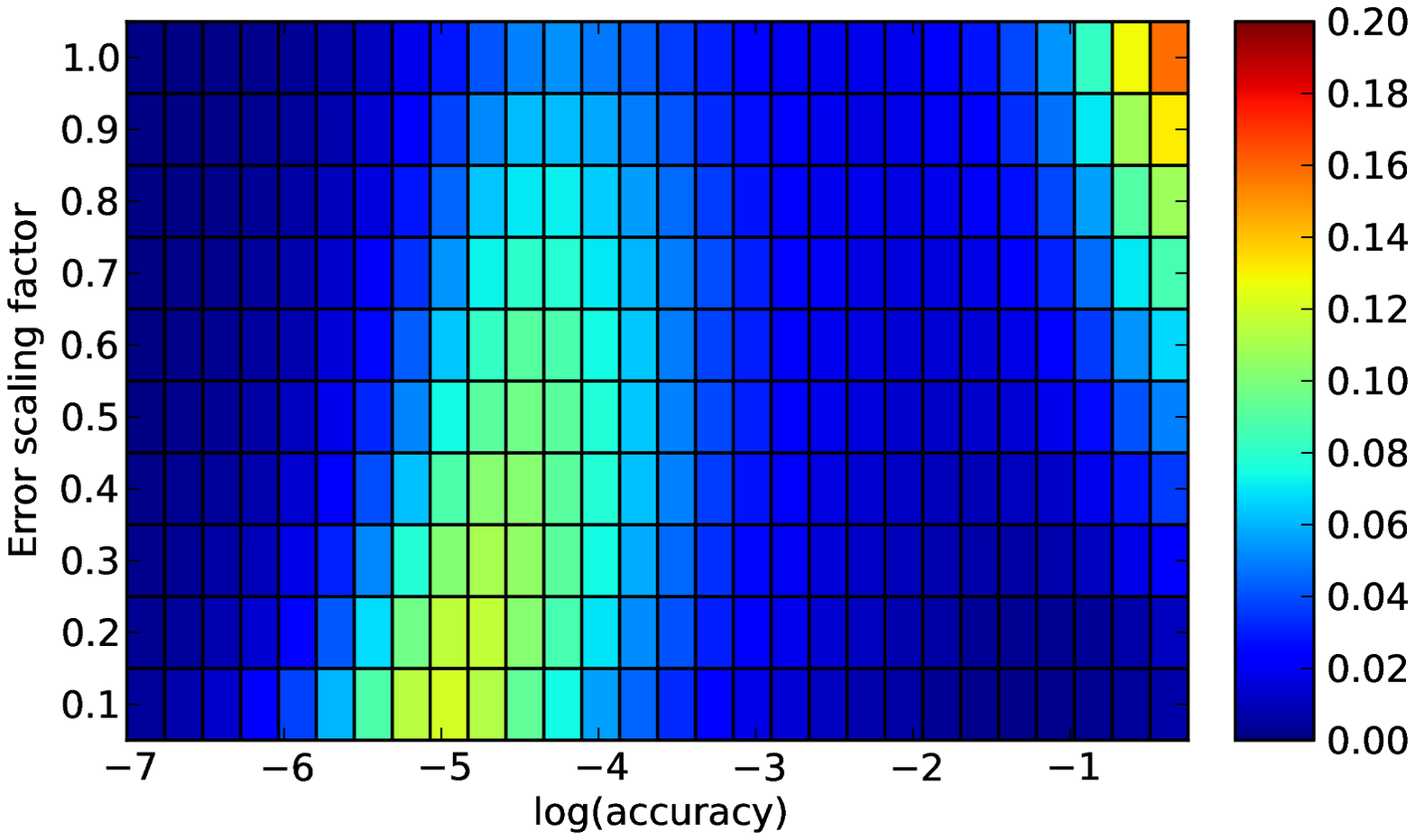} & 
\includegraphics[width=3.45in]{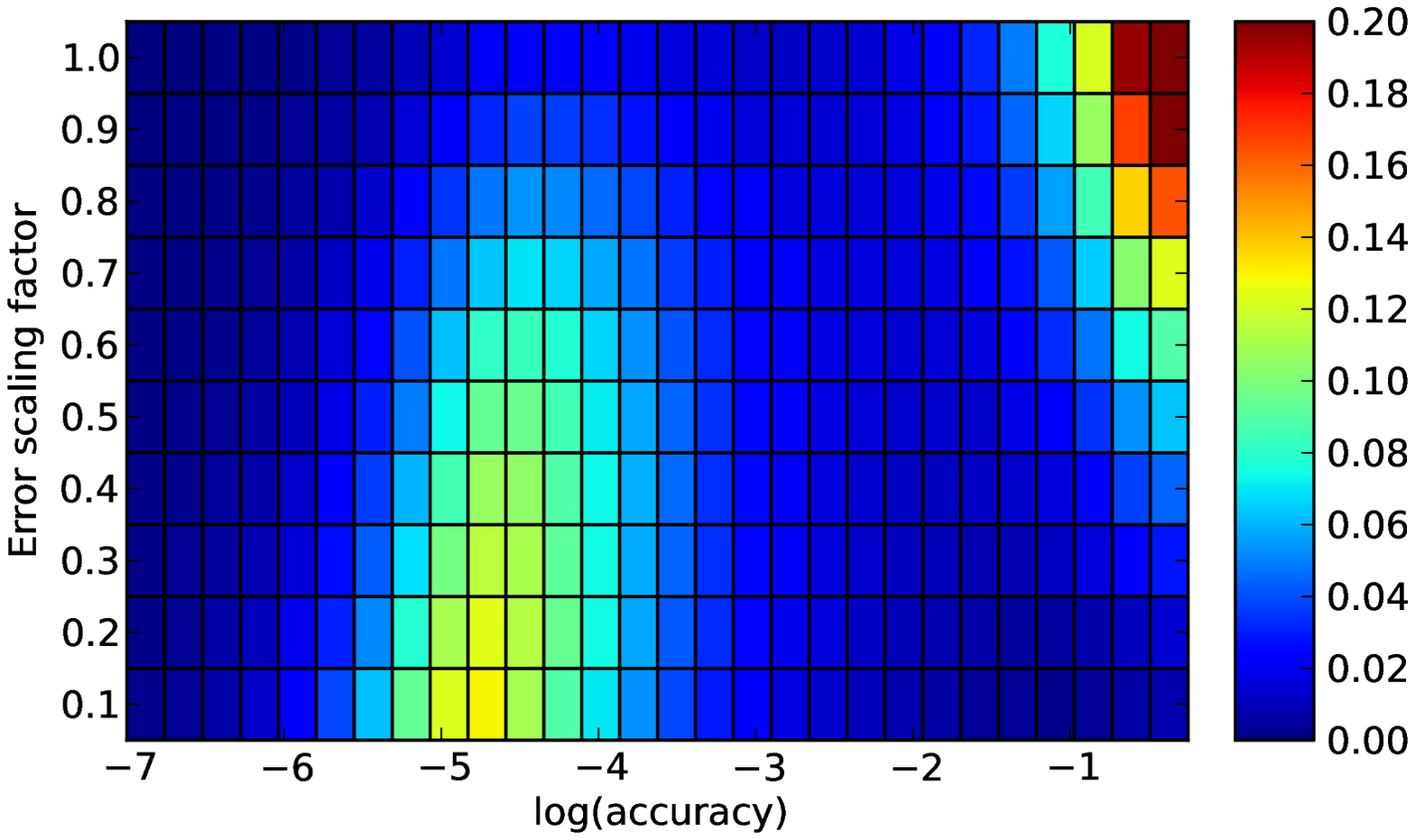} \\
\end{tabular}
\end{figure*}

Fig.~\ref{err} shows the overall accuracy distributions for the two entropy methods for the different values of the error scaling factor used. Again both methods show a dependency on how noisy the light curve is but the conditional entropy performs slightly better in all cases, i.e., for a particular accuracy cutoff value, the conditional entropy returns a larger number of periods than the Shannon entropy. This also shows the harmonic data with the conditional entropy method a much better indicator of periodicity for all noise levels. Note that for $B > 0.7$, most of the Shannon entropy accuracies are significantly wrong (the strong concentration in the top right hand corner of Fig.~\ref{err}(d)).

Although we have included random sampling and a noise term in our generated data, we have so far only 
demonstrated the efficacy of the algorithm with a synthetic sinusoidal signal which is not the most realistic situation. However, as Table~\ref{table1} shows, when applied to real data with all its additional characteristics (such as observing cadences rather than random sampling and heteroscedastic errors), the conditional entropy method is vastly more effective and robust. We note that \cite{ckp} get fractional recovery rates of 0.88 and 0.99 for the true period and an integer (sub)multiple of the period respectively for an accuracy cutoff of $5 \times 10^{-3}$. However, we reserve a far more extensive comparison of the conditional entropy algorithm to other period finding techniques with real data to our companion paper (\cite{graham13})

\begin{figure}
\caption{This shows the distribution of accuracies for the MACHO data in terms of the number of observations per cycle for the two entropy-based methods: red (conditional entropy) and blue (Shannon entropy). }
\label{tscnpt}
\includegraphics[width=3.45in]{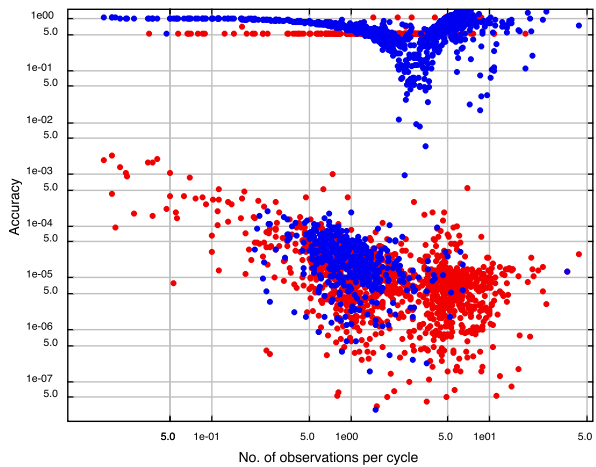}
\end{figure}

The accuracy distributions for the two entropy-based methods are shown in Fig.~\ref{tscnpt}. The line of CE points at log(accuracy) = 0.5 indicates those light curves (12\%) for which the method has incorrectly recovered a half period. As expected, these are predominantly eclipsing binaries with a few RRCs as well.
A large fraction of the Shannon entropy periods (blue points) are clearly around the $\sim1$ day value (the phenomenon shown in Fig.~\ref{solar}). In fact, this is also class-related behaviour with the Shannon entropy method only correctly recovering the true periods for mainly Cepheid variables. Of the three classes in this data set, the distinguishing feature of the Cepheids is that they have a higher S/N than RR Lyrae or eclipsing binaries. This makes it easier for the Shannon entropy method to identify their correctly phased light curves easier than the other two classes. Again we present a more extensive discussion of the class dependencies of various period-finding algorithms in our companion paper.

\begin{table}
\caption{This shows the fractional recovery rate of true periods for the two entropy algorithms with the real MACHO data and different accuracy cutoffs.}
\label{table1}
\begin{tabular}{lllllll}
\hline
Method & \multicolumn{3}{c}{True period} & \multicolumn{3}{c}{Harmonic} \\
 & $10^{-5}$ & $10^{-4}$ & $10^{-3}$ & $10^{-5}$ & $10^{-4}$ & $10^{-3}$ \\
\hline
Conditional & 0.47 & 0.82 & 0.86 & 0.52 & 0.94 & 0.99 \\ 
Shannon  & 0.07 & 0.28 & 0.29 & 0.07 & 0.29 & 0.30\\
\hline
\end{tabular}
\end{table}

\section{Conclusions}
In this paper we have introduced a new period finding algorithm based on the conditional entropy of the time series. As \cite{cincotta99} suggested, this improves on the results of a basic Shannon entropy-based approach. Although conditional entropy shows similar results to existing algorithms when applied to standard synthetic data, it shows itself to be much a more powerful technique with detecting general periodicity (via an integer (sub)multiple of the true period) and real data. This stresses the importance of using real data whenever possible to test new techniques. Although we have only considered the application of this algorithm to single-band light curves, we think that the technique can easily be extended to multiband light curves and also to transit searches and will explore these in a future paper. 

\section*{Acknowledgments}

We thank the referee, Pablo Cincotta, for his useful comments.

This work was supported in part by the NSF grants AST-0909182 and IIS-1118041, by the W. M. Keck Institute for Space Studies, and by the U.S. Virtual Astronomical Observatory, itself supported by the NSF grant AST-0834235.

\appendix

\section[]{Optimal estimator for the conditional entropy}
\label{appa}

 \cite{bb} describes an algorithm --  Bayesian Blocks -- that finds the optimal segmentation of 1D data in an observation interval. This can be extended to arbitrary dimensions in the following way (Scargle, private communication): 

\begin{enumerate}
\item compute the 2D Voronoi tessellation of the points   
\item compute the areas of the Voronoi cells
\item sort the (1D array of) areas (increasing)
\item feed this array to the 1D Bayesian Block algorithm
\item the blocks coming out of the previous step will in general be broken up into non-connecting pieces, so
at this point it may be necessary to identify these pieces -- yielding a set of blocks (hyper-Voronoi regions) that are connected subsets of the Voronoi cells  in the original blocks.
\end{enumerate}

The (Shannon) entropy of the point distribution (light curve) can be estimated (\cite{miller03}) as: 
\[
H_V = \sum_{i=1}^{m} \frac{C(U^i)}{N} \log \left( \frac{N A(U^i)}{C(U^i)} \right) \]

\noindent
where each hyper-Voronoi region $U^i$ has $C(U^i)$ Voronoi regions in it, $N = \sum_i C(U^i)$, and $A(U^i)$ is the D-dimensional volume of $U^{i}$. The conditional entropy is then given by  $H(m|\phi) = H(m, \phi) - H(\phi)$ where $H_V = H(m, \phi)$. $H(\phi)$ can be estimated from the Bayesian Block (BB) partitioning of the phase distribution via: $H(\phi) = - \sum_{i=1}^{m} f(x_i) \log (f(x_i) / w(x_i))$ where $f(x_i)$ is the fraction of points in the $i^{\mathrm{th}}$ BB partition and $w(x_i)$ is its width.

\label{lastpage}


\begin{thebibliography}{99}
\bibitem[\protect\citeauthoryear{Alcock et al.}{2003}]{macho}Alcock C., et al., 2003, Variable Stars in the Large Magellanic Clouds (MACHO, 2001)
\bibitem[\protect\citeauthoryear{Baluev}{2012}]{baluev12} Baluev R.~V., 2012, in Fifty years of Cosmic Era: Real and Virtual Studies of the Sky, Mickaelian A.~M., Malkov O. Yu., Samus N.~N., eds. NAS RA, Yerevan, 230
\bibitem[\protect\citeauthoryear{Baluev}{2013}]{baluev} Baluev R.~V., 2013, MNRAS, in press (arXiv:1302.1068)
\bibitem[\protect\citeauthoryear{Cincotta et al.}{1995}]{cnm95} Cincotta P.~M. Mendez M., Nunez J.~A., 1995, ApJ, 449, 231
\bibitem[\protect\citeauthoryear{Cincotta et al.}{1999}]{chmn99} Cincotta P.~M., Helmi A., Mendez M., Nunez J.~A., Vucetich H., 1999, MNRAS, 302, 582
\bibitem[\protect\citeauthoryear{Cincotta}{1999}]{cincotta99} Cincotta P.~M., 1999, MNRAS, 307, 941
\bibitem[\protect\citeauthoryear{de Jager, Raubenheimer \& Swanepoel}{1989}]{dejager} de Jager O.~C., Raubenheimer B.~C., Swanepoel J.~W.~H., 1989, A\&A, 221, 180
\bibitem[\protect\citeauthoryear{Drake et al.}{2009}]{crts} Drake A.~J., et al., 2009, ApJ, 696, 870
\bibitem[\protect\citeauthoryear{Drake et al.}{2013}]{rrlyrae} Drake A.~J., et al., 2013, ApJ, 763, 1
\bibitem[\protect\citeauthoryear{Foster}{1996}]{foster} Foster G., 1996, AJ, 112, 1709
\bibitem[\protect\citeauthoryear{Graham et al.}{2013}]{graham13} Graham M.~J., Drake A.~J., Djorgovski S.~G., Mahabal A.~A., Donalek C., Duan V., Maher A., 2013, MNRAS, submitted 
\bibitem[\protect\citeauthoryear{Gregory \& Loredo}{1992}]{gregory92} Gregory P.~C., Loredo T.~J., 1992, ApJ, 398, 146
\bibitem[\protect\citeauthoryear{Hogg}{2008}]{hogg08} Hogg D., 2008, http://arxiv.org/abs/0807.4820
\bibitem[\protect\citeauthoryear{Huijse et al.}{2011}]{correntropy} Huijse P., Estevez P.~A., Zegers P., Principe J.C., Protopapas P., 2011, IEEE Signal Processing Letters, 18 (6), 371
\bibitem[\protect\citeauthoryear{Huijse et al.}{2012}]{ckp} Huijse P., Estevez P.~A., Protopapas P., Zegers P., Principe J.~C., 2012, IEEE Trans. Signal Processing, 60, 10, 5135
\bibitem[\protect\citeauthoryear{Hutchinson  \& de Hoog}{1985}]{gcvs85} Hutchinson M.~F., de Hoog F.~R., 1985, Numer. Math, 47, 99
\bibitem[\protect\citeauthoryear{Ivezic et al.}{2011}]{lsst} Ivezic Z., et al., 2011, arXiv:0805.2366
\bibitem[\protect\citeauthoryear{Kaiser et al.}{2004}]{panstarrs} Kaiser, N., et al., 2002, SPIE 4836, 154
\bibitem[\protect\citeauthoryear{Kovacs}{1980}]{kovacs} Kovacs G., 1980, Ap\&SS, 69, 485
\bibitem[\protect\citeauthoryear{Kato \& Uemura}{2012}]{lasso} Kato T., Uemura M., 2012, PASJ, (arXiv:1205.4791)
\bibitem[\protect\citeauthoryear{Leroy}{2012}]{fastfourier} Leroy B., 2012, A\&A, 545, 50
\bibitem[\protect\citeauthoryear{Lomb}{1976}]{lomb} Lomb N.~R., 1976, Ap\&SS, 39, 447
\bibitem[\protect\citeauthoryear{Loredo}{2012}]{loredo11} Loredo T., 2011, in New Horizons in Time Domain Astronomy, Proc. IAU Symp. No. 285, Griffin E., Hanisch R., \& Seaman R., eds. Cambridge University Press, Cambridge. http://arxiv.org/abs/1201.4114
\bibitem[\protect\citeauthoryear{Miller}{2003}]{miller03} Miller E.~G., 2003, IEEE International Conference on Acoustics, Speech, and Signal Processing, 3
\bibitem[\protect\citeauthoryear{Misha et al.}{2011}]{misha11} Misha B.~P., Principe J.~C., Estevez P.~A., Protopapas P., 2011, IEEE International Workshop on Machine Learning for Signal Processing
\bibitem[\protect\citeauthoryear{Pojmanski}{2002}]{asas} Pojmanski G,. 2002, Acta Astron., 52, 397
\bibitem[\protect\citeauthoryear{Pojmanski et al.}{2005}]{acvs} Pojmanski G., Pilecki B., \& Szczygiel D., 2005, Acta Astron., 55, 275
\bibitem[\protect\citeauthoryear{Rau et al.}{2009}]{ptf} Rau A., et al., 2009, PASP, 121, 1334
\bibitem[\protect\citeauthoryear{Richards et al.}{2012}]{macc12} Richards J.~W., Starr D.~L., Miller A.~A., Bloom J.~S., Butler N.~R., Brink H., Crellin-Quick A., 2012, ApJ, submitted (arXiv:1204.4180)
\bibitem[\protect\citeauthoryear{Scargle}{1982}]{scargle} Scargle J.~D., 1982, ApJ, 263, 835
\bibitem[\protect\citeauthoryear{Scargle et al.}{2012}]{bb} Scargle J.~D., Norris J.~P., Jackson B., Chiang J., 2012, ApJ, submitted (arXiv:1207.5578)
\bibitem[\protect\citeauthoryear{Schwarzenberg-Czerny}{1989}]{aov89} Schwarzenberg-Czerny A., 1989, MNRAS, 241, 153
\bibitem[\protect\citeauthoryear{Schwarzenberg-Czerny}{1999}]{schwarzenberg99} Schwarzenberg-Czerny A., 1999, ApJ, 516, 315
\bibitem[\protect\citeauthoryear{Stellingwerf}{1978}]{stellingwerf} Stellingwerf R.~F., 1978, ApJ, 224, 953
\bibitem[\protect\citeauthoryear{Stellingwerf}{2011}]{pdm2} Stellingwerf R.~F., 2011, in RR Lyrae Stars, Metal-Poor Stars, and the Galaxy, Carnegie Observatories Astrophysics Series Vol. 5, McWilliam A., ed. Observatories of the Carnegie Institution of Washington, Pasadena, 47
\bibitem[\protect\citeauthoryear{Wang, Khardon \& Protopapas}{2012}]{wang12} Wang Y., Khardon R., Protopapas P., 2012, ApJ, 756, 67
\bibitem[\protect\citeauthoryear{Zechmeister \& Kurster}{2009}]{gls} Zechmeister M., Kurster M., 2009, A\&A, 496, 577

\end{thebibliography}
\end{document}